\documentclass[doublecol]{epl2} 
\begin{document}
\title{Driven polymer translocation through a nanopore: \\a manifestation of
anomalous diffusion}
%\shorttitle{Driven polymer translocation through a nanopore} 
%Insert here a short version of the title if it exceeds 70 characters

\author{J. L. A. Dubbeldam \inst{1,2} \and  A. Milchev \inst{1,3} ,  V.G.
Rostiashvili \inst{1} \and T.A. Vilgis \inst{1}}

\institute{                    
  \inst{1} Max Planck Institute for Polymer Research, 10 Ackermannweg 55128
Mainz, Germany\\
  \inst{2}  Delft University of Technology 2628CD Delft, The Netherlands\\
   \inst{3} Institute for Physical Chemistry Bulgarian Academy of Science, 1113
Sofia, Bulgaria
}

\abstract{We study the translocation dynamics of a polymer chain threaded
through a nanopore by an external force. By means of diverse methods (scaling
arguments, fractional calculus and Monte Carlo simulation) we show that the
relevant dynamic variable, the translocated number of segments $s(t)$, displays
an {\em anomalous} diffusive behavior even in the {\em presence} of an external force. The 
anomalous dynamics of the translocation process is governed by the same universal
exponent $\alpha = 2/(2\nu +2 - \gamma_1)$, where $\nu$ is the Flory exponent and 
$\gamma_1$ - the surface exponent, which was established recently for
the case of non-driven polymer chain threading through a nanopore.
A closed analytic expression for the probability distribution function $W(s, t)$, 
which follows from the relevant {\em fractional} Fokker - Planck equation, 
is derived in terms of the polymer chain length $N$ and the applied drag force $f$. 
It is found that the average translocation time scales as $\tau \propto f^{-1}N^{\frac{2}{\alpha} -1}$. 
Also the corresponding time dependent statistical moments, $\left\langle s(t)
\right\rangle \propto t^{\alpha}$ and $\left\langle s(t)^2 \right\rangle \propto
t^{2\alpha}$ reveal unambiguously the anomalous nature of the translocation dynamics 
and permit direct measurement of $\alpha$ in experiments. These findings are tested
and found to be in perfect agreement with extensive Monte Carlo (MC) simulations.
}

\maketitle

\section{Introduction} Recently single molecule experiments probing single -
stranded DNA or RNA translocation through a membrane nanopore attracted
widespread attention \cite{Meller}. These investigations have been triggered in 
the seminal experimental paper by Kasianowicz {\it et al.} \cite{Kasianowicz}
where an electric field drives single - stranded DNA and RNA molecules through
the $\alpha$ - hemolysin nanopore so that  each  threading  is signaled by the
blockage of the ion current. By recording the blockage time one can reconstruct
the whole driven translocation of DNA molecule. More recently solid - state
nanopores have been used for DNA translocation experiment \cite{Li,Storm}. Such
pores can be tuned in size and are more stable over a wide range of voltages,
temperature as well as the solvent pH.

The physical nature of the translocation process is still not well understood.
The theoretical consideration of the translocation dynamics is usually based on
the assumption that the translocation length $s$ (i.e. the translocated number
of segments at time $t$) is the only relevant dynamic variable which is governed
by a conventional Brownian diffusion process \cite{Sung1,Sung2,Muthu}. The
main predictions for the average translocation time $\tau$ looks as follows. For
an unbiased translocation $\tau (N) \propto a^2 N^2/D$ (here $a$ is a polymer
Kuhn segment length and $D$ is a diffusion coefficient whose 
N-dependence is not well established) whereas the
$\tau$ for the driven translocation (when a  polymer experiences a chemical
potential difference $\Delta\mu$  between the environments separated by the
membrane) scaled as $\tau \propto T a^2 N/(D \Delta\mu) $. Here $T$ denotes temperature
and we have set the Boltzmann coefficient $k_B \equiv 1$. More recently Kantor
\& Kardar \cite{Kantor1,Kantor2}have cast doubt on these results by noting that
the unimpeded motion of a polymer scales as the characteristic Rouse time
$\tau_{\rm Rouse} \propto N^{2\nu+1}$ where $\nu = 0.588$ at $d = 3$ \cite{Doi},
so that $\tau_{\rm Rouse} \gg \tau$ although the unimpeded motion must be faster
than that of a constrained chain. Kantor \& Kardar argue
\cite{Kantor1,Kantor2} that the {\it lower bound } of the translocation time
(which corresponds to an unimpeded motion of a polymer) should scale as $\tau
(\Delta\mu = 0) \propto N^{2\nu+1}$ and $\tau (\Delta\mu) \propto
N^{1+\nu}/\Delta\mu$ for an unbiased  and driven translocation cases
correspondingly. They also carried out MC - simulations (using a bond
fluctuating model on a $2d$ lattice) and the results show that $\tau (\Delta\mu = 0)
\propto N^{2.5}$ and $\tau (\Delta\mu) \propto N^{1.53}/\Delta\mu$, i.e. at least in the
case of the driven translocation the theory is inconsistent with MC - simulation
data. A more recent MC and Langevin dynamics study \cite{Ala-Nissila} reports
scaling laws  $\tau \propto N^{1.5}$ and $\tau \propto N^{1.65}$ for
relatively short and relatively long chains respectively. It has also been
questioned \cite{Kantor1,Kantor2} whether the translocation dynamics is that of 
normal Brownian motion and suggested instead that {\it anomalous diffusion
dynamics}
\cite{Metzler1} might explain some MC - findings. Nevertheless, there is so far
no clear knowledge  regarding the physical origin of such anomalous dynamics.
It is also not clear how one can make use of the {\it fractional} Fokker - Planck equation
\cite{Metzler1,Metzler2} which seems to govern this type of dynamics.

In this paper we come up with a general picture of the driven polymer
translocation based on our previous consideration of the unbiased problem
\cite{Dubb}. We first sketch the mapping of the $3d$ problem on the $1d$
translocation $s$ - coordinate. This leads to an anomalous diffusion in the
external force field which one could quantify in terms of the {\em fractional} Fokker -
Planck equation (FFPE). The solution of this equation is then obtained on the
interval $0 \leq s \leq N$ in a closed analytical form . The subsequent
comparison of our extensive MC - results with the proper analytical expressions
shows a very nice quantitative agreement.

\section{Dynamics in terms of a single translocation coordinate} 
\begin{figure}[h]
%\twofigures[scale=0.3]{sketch.eps}{barrier.eps}
\centerline{\includegraphics[scale=0.3]{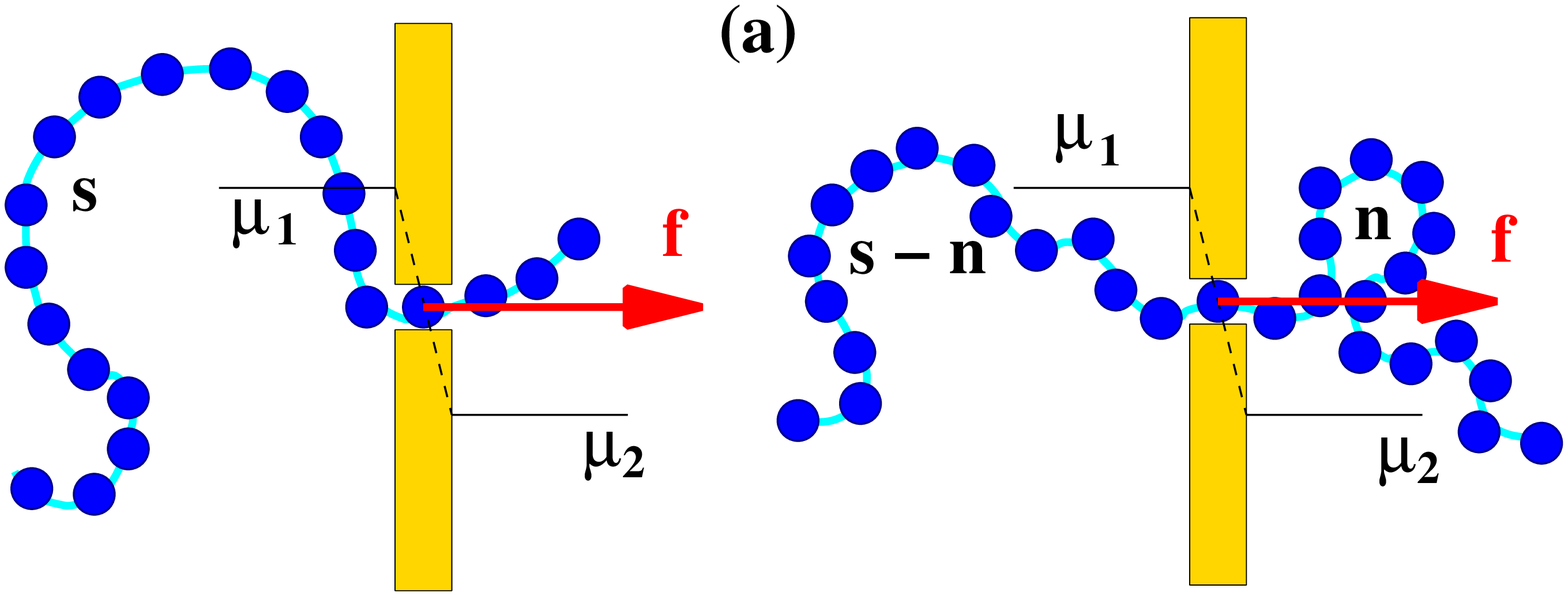}}
\centerline{\includegraphics[scale=0.3, angle=270]{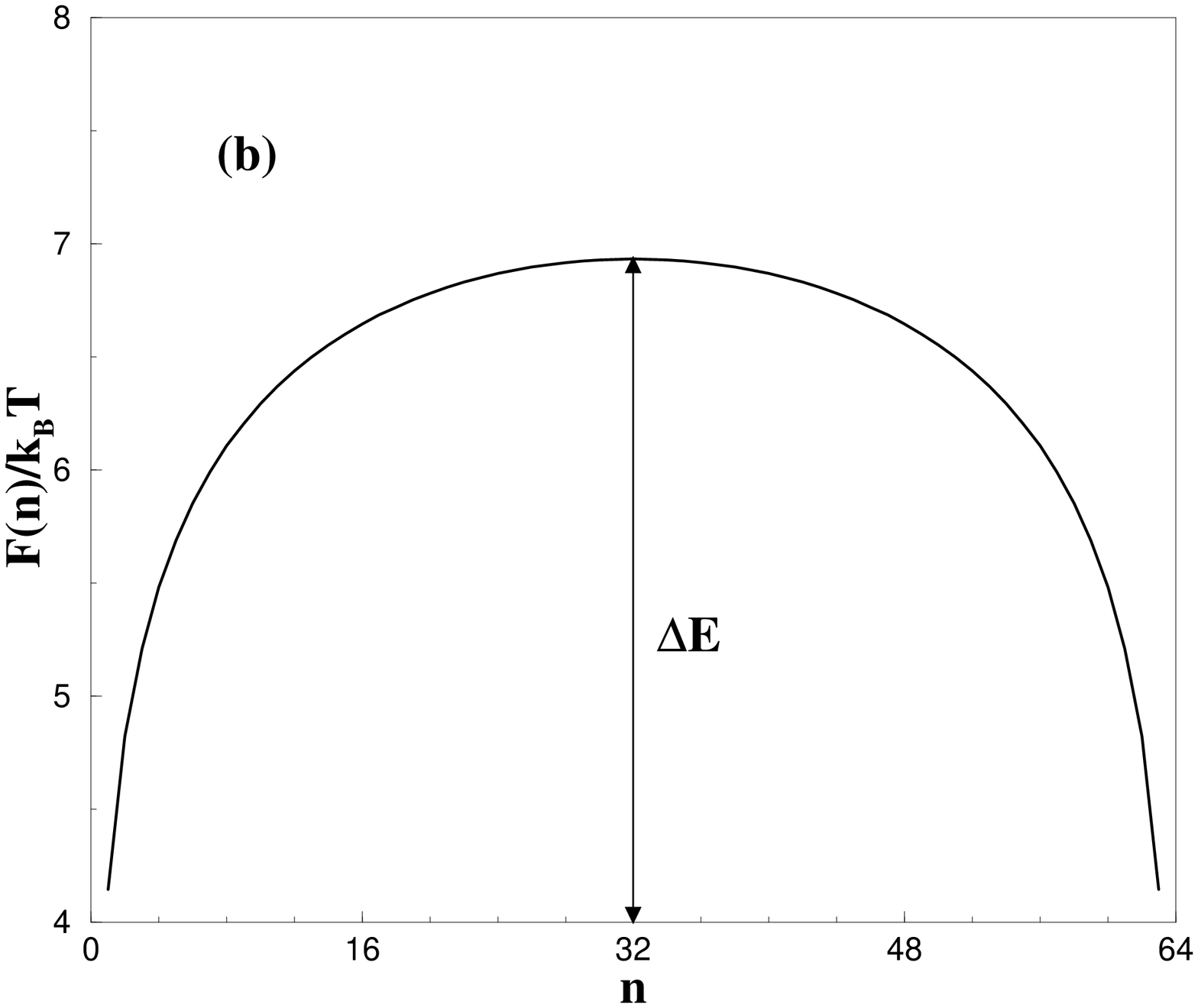}}
\caption{How a fold squeezes through a nanopore. The driving force $f$ is caused
by a chemical potential gradient $\Delta\mu = \mu_1 - \mu_2$. (a) The fold of
the length $s$ is fragmented into $n$ and $s - n$ - parts during its threading.
(b) This fragmentation gives rise to an effective entropic barrier $F (n)$
with height $\Delta E$ at $n=s/2$.  }
\label{Sketch}
%\twocolumn
\end{figure}

As already noted, the
initial $3d$ problem can be rephrased in terms of $1d$ translocation 
coordinate $s$, and in doing so one arrives at a typical case of anomalous diffusion. 
Recently we suggested \cite{Dubb} that the translocation proceeds by successive threading
of small fractions of the polymer, called {\em folds}, which equilibrate fast enough
compared to the whole chain, and can be considered as building blocks of such mapping. 
In a somewhat different context concerning the polymer dynamics the notion of folds 
has been discussed earlier \cite{Shura}. 
Figure \ref{Sketch} shows how a fold overcomes an entropic barrier 
caused by a narrow pore.
%\onecolumn

If the fold is fragmented into $n$ and $s - n$ parts while it is threading
through the pore then the corresponding free energy reads $F (n)/T =  - s\ln
\kappa - (\gamma_1  - 1) \ln [n(s-n)]$, where $\kappa$ is the connective
constant and $\gamma_1$ is the surface entropic exponent \cite{Vander}. Then the
corresponding activation barrier which could be associated with the fold
threading can be calculated as $\Delta E (s) = F (s/2) -  F  (1) = (1 -
\gamma_1) T \ln s$.

In the force - free case the characteristic time of the fold transition from 
{\it cis -} to the {\it trans -} side of the membrane can be estimated as follows.
In the absence of a separating membrane this would be the pure
Rouse time $t_{\rm Rouse} \propto s^{2\nu + 1}$. The membrane with a nanopore
imposes
an additional entropic activation barrier $\Delta E (s)$ which slows down the
transition rate.
The characteristic time, therefore, scales as $t(s) = t_{\rm Rouse} (s) \exp
[\Delta E (s)] \propto s^{2\nu + 2 - \gamma_1}$. This makes it possible to
estimate the mean-squared displacement of the $s$ - coordinate as
\begin{equation}
\left\langle s^2 \right\rangle \propto t^{2/(2\nu + 2 - \gamma_1)} .
\end{equation}
As a result the mapping on the $s$ coordinate leads to an {\em anomalous
diffusion} law, $\left\langle s^2\right\rangle \propto t^{\alpha}$, where
$\alpha = 2/(2\nu + 2 - \gamma_1)$. Taking into account that  for $d = 3$ , $\nu
= 0.588$ and $\gamma_1 = 0.680$ \cite{Diehla}, we obtain  $\alpha = 0.801$. In
turn, the average translocation time $\tau \propto N^{2/\alpha} \propto
N^{2.496}$. Remarkably, in $2 d$ where $\nu_{2d}=0.75$ and $\gamma_1\approx
0.945$\cite{Whittington},one finds $\alpha\approx 0.783$, i.e. $\alpha$ is
almost dimensionality independent!  This explains why the measured exponents in
both $2d$ \cite{Kantor1} and in $3d$ \cite{Barkema}  are so close. The presence
of the external force imposed on the translocating chain leads to a nonisotropic
{\it cis - trans} - transition of the folds. It can be quantified within the FFPE 
- formalism which was originally suggested by Barkai, Metzler and Klafter 
\cite {Barkai}.

\section{Fractional Fokker - Planck equation}  The formalism of FFPE provides an
appropriate technique which describes the anomalous diffusion in an external
force - field. In our case FFPE has the form
\begin{equation}
\frac{\partial}{\partial t} \:  W (s, t) = {_0}D_{t}^{1 - \alpha}\left[
\frac{\partial}{\partial s} \frac{U'(s)}{\xi_{\alpha}}+ K_{\alpha}
\frac{\partial^2}{\partial s^2} \right] \: W (s, t) ,
\label{FFDE}
\end{equation}
where $W (s, t)$ is the probability distribution function (PDF) for having a
segment $s$ at time $t$ in the pore, and the fractional Riemann - Liouville
operator  ${_0}D_{t}^{1 - \alpha} W (s, t) = (1/\Gamma (\alpha))
(\partial/\partial t)
\int_0^t d t' W (s, t')/(t - t')^{1-\alpha}$. In Eq. (\ref{FFDE}) $\Gamma
(\alpha) $ is the Gamma-function, $K_{\alpha}$ is the so called generalized
diffusion constant, and $\xi_{\alpha}=T/K_{\alpha}$ is the appropriate friction 
coefficient. In our case the external field $U (s)$ is a simple linear
function of the translocation $s$ - coordinate, namely $U (s) = - \Delta\mu \:
s$, where $\Delta\mu = \mu_1 - \mu_2$. We consider the boundary value problem
for FFPE \cite{RMJK} in the interval $0\leq s \leq N$. The boundary conditions correspond to
the {\em reflecting-adsorbing} case, i.e., $[U'(s) W(s, t)/T + (\partial/\partial s) W (s,
t)]|_{s=0} = 0$ and $W (s = N, t) = 0$. The initial distribution is concentrated
in $s_0$, i.e., $W (s, t=0) = \delta (s - s_0)$. The full solution can be
represented as a sum over all eigenfunctions $\psi_n (s)$ \cite{Risken}, i.e.,
$W (s, t) = \exp(\Phi(s_0)-\Phi(s)) \sum_{n=0}^{\infty} T_n (t) \psi_n
(s)\psi_n(s_0)$ where $\Phi(s)=U(s)/2T$ , $\psi_n(s)=\exp(\Phi(s))\varphi_n (s)$
and $\varphi_n (s)$ obey the equations $\left[  (d^2/ds^2) - f \: (d / ds )  +
\lambda_{n,\alpha}/K_{\alpha} \right] \varphi_n (s) = 0 $ (where $f \equiv
\Delta\mu/T$) , and the eigenvalues $\lambda_{n,\alpha}$ can be readily found
from the foregoing boundary conditions. The temporal  part $T_n (t)$ obeys the
equation $(d/d t) T_n (t) = - \lambda_{n,\alpha} \: {_0}D_{t}^{1 - \alpha} T_n
(t)$. The solution of this equation is given by $T_n (t) = T_n (t=0) E_{\alpha}
(- \lambda_{n,\alpha} \: t^{\alpha})$ \cite{Metzler1} where the Mittag - Leffler
function $E_{\alpha}(x)$ is defined by the series expansion $E_{\alpha} (x) =
\sum_{n=0}^{\infty} \: x^k/\Gamma (1+\alpha k)$. At $\alpha = 1$ it turns back
into a standard exponential function (normal diffusion). Allowing for the boundary
conditions leads to a transcendental equation for the eigenvalues, i.e.
$-2\sqrt{\kappa_n}/f = \tan (\sqrt{\kappa_n} N)$, where $\kappa_n =
\lambda_{n,\alpha}/K_{\alpha} - f^2/4$. This eigenvalue problem has 
simple solutions in two limiting cases. For a very weak force $f N \ll 1$ the
$\kappa_n = \lambda_{n, \alpha}/K_{\alpha} =(2n + 1)^2 \pi^2/4N^2$ and the
eigenfunctions take on the form $\varphi_n (s) = \sqrt{2/N} \cos \left[ (2n + 1)\pi
s/2N\right] $. The resulting solution for $W (s,t)$  at $f=0$ reduces to that
of the force - free case \cite{Dubb}.

In this paper we focus our attention on the opposite limit, $f N \gg 1$, i.e.
when the driving force is pretty strong. In this case the eigenvalues spectrum
reads $\lambda_{n, \alpha} = (f^2/4 + n^2\pi^2/N^2) K_{\alpha}$ and the
eigenfunctions $\psi_n (s) = \sqrt{2/N} \sin (n \pi s/N)$, so that the resulting
solution becomes
\begin{eqnarray}
W (s, t)&=&\frac{2}{N} \: {\rm e}^{f (s - s_0)/2}\:  \sum_{n = 0}^{\infty}
\:\sin\left[ \frac{n
\pi s_0}{N}\right] \: \sin\left[ \frac{n \pi s}{N}\right]
\nonumber\\
&\times& E_{\alpha}\left[ - \left( \frac{f^2}{4} + \frac{n^2 \pi^2}{N^2}\right)
\: K_{\alpha} \:
t^{\alpha}\right] \quad.
\label{Solution_box}
\end{eqnarray}

In the limit of strong driving force the translocation times are relatively (as
compared to the force - free case)  short and we could use the small argument
approximation for the  Mittag - Leffler function $ E_{\alpha} (-x)$, i.e.
$E_{\alpha} (- x) \simeq  \exp [- x/\Gamma(1+\alpha)]$ at $x \ll 1$. This
makes it possible to obtain an explicit analytical expression for $W(s, t)$ which  can be
derived by replacing  the summation by an integral in eq.
(\ref{Solution_box}). In doing so one should use the relation $2 \sin(n \pi
s_0/N)\sin(n \pi s/N) = \cos [n \pi (s-s_0)/N ] - \cos [n \pi (s + s_0)/N]$.
Then one can integrate over $n$ explicitly, taking  the limit $s_0 \rightarrow 0$, and
finally normalize the FPTD: $w (s, t) \equiv \lim_{s_0 \rightarrow 0} W(s,
t)/\int_{0}^{N} W(s, t) ds $.  This yields eventually
\begin{eqnarray}
w (s, t) = \frac{\exp\left\lbrace -(s - f {\tilde t})^2/4 {\tilde
t}\right\rbrace }{\sqrt{\pi {\tilde t}}\left\lbrace {\rm erf}[f \sqrt{{\tilde
t}}/2] - {\rm erf} [ (f {\tilde t} - N)/2\sqrt{{\tilde t}}] \right\rbrace },
\label{Norma}
\end{eqnarray}
where the dimensionless force $f = \Delta\mu/T$ ,  ${\tilde t} =  K_{\alpha}
t^{\alpha}/\Gamma(1+\alpha)$ and ${\rm erf} (x)$ is the error function.
Our further theoretical findings are based mainly on eqs. (\ref{Solution_box})
and (\ref{Norma}) for  PDF.

\section{First - passage time distribution} In the chain translocation
experiment the initial position $s_0$ can be  fixed and the distribution of the
translocation times is actually equivalent to the {\it first - passage time
distribution} (FPTD) $Q (s_0, t)$ \cite{Risken}.  The relation $Q (s_0, t) = -
(d/dt) \int_{0}^N W(s, t) ds$ \cite{Risken} enables to calculate FPTD
explicitly. Starting from eq. (\ref{Solution_box}) we arrive at the expression
\begin{eqnarray}
Q(s_0, t)&=&\frac{\pi K_{\alpha}{\rm e}^{f (N - s_0)/2}}{N^2t^{1-\alpha}}  \:
\sum_{n = 0}^{\infty}
\:(-1)^{(n-1)} \sin\left( \frac{n\pi s_0}{N}\right)
\nonumber\\
 &\times& E_{\alpha, \alpha}\left[ - \left( \frac{f^2}{4} + \frac{n^2
\pi^2}{N^2}\right) \:
K_{\alpha} \:  t^{\alpha}\right] \quad,
\label{Final_time}
\end{eqnarray}
where the generalized Mittag - Leffler function $ E_{\alpha, \alpha} (x) = 
\sum_{k = 0}^{\infty} \: x^k/\Gamma (\alpha + k \alpha)$.

In the same manner as above  we could use the small argument approximation for
the generalized Mittag - Leffler function $ E_{\alpha, \alpha} (-x)$, i.e.
$E_{\alpha, \alpha} (- x) \simeq (\alpha/\Gamma(1+\alpha)) \exp [-
x/\Gamma(1+\alpha)]$ at $x \ll 1$, to obtain the explicit analytical expression
for $Q(s_0, t)$.  The  substitution of the summation by integration in eq.
(\ref{Final_time}) and the  use of the relation $(-1)^{n-1} \sin(n\pi s_0/N) =
-\cos(n\pi) \sin(n\pi s_0/N)=\sin[n\pi(1-s_0/N)]-\sin[n\pi(1+s_0/N)]$ enable 
finally to obtain for the  normalized FPTD, $\lim_{s_0 \rightarrow 0} Q(s_0,
t)/\int Q(s_0, t) dt \rightarrow Q (t)$, the following expression
\begin{eqnarray}
Q(t) &=& \frac{\alpha}{4\pi^{1/2} f t} \left[ \frac{\Gamma(1+\alpha)}{K_{\alpha}
t^{\alpha}}\right]^{1/2}\left[ \frac{N^2\Gamma(1+\alpha)}{K_{\alpha} t^{\alpha}}
- 2\right] 
\nonumber\\
&\times& \exp \left\lbrace -\frac{\left[ N - f \frac{K_{\alpha}
t^{\alpha}}{\Gamma(1+\alpha)}\right]^2 }{4\frac{K_{\alpha}
t^{\alpha}}{\Gamma(1+\alpha)}} \right\rbrace .
\label{FPDT_norma}
\end{eqnarray}
As one can see, after normalization the dependence on the initial value $s_0 \rightarrow 0$
drops out. It is of interest that FPDT given by eq. (\ref{FPDT_norma}) exactly
coincides (at $\alpha = 1$, i.e., in the Brownian dynamics limit)  with the
corresponding expression in the paper by Lubensky \& Nelson \cite{Lubensky}. It
is also evident from eq. (\ref{FPDT_norma}) that the maximum position scales
as $t_{\rm max} \propto (N/f)^{1/\alpha} = (N/f)^{1.25}$. Nevertheless, the
function $Q (t)$ is quite skewed and we will see below that the average
translocation time $\tau = \int t Q (t) dt$ (which is presumably measured in an
experiment) scales differently. Note that Eq. (\ref{FPDT_norma}) is valid for 
$t \le N^{2/{\alpha}}\left(\Gamma(1+\alpha)/2K_\alpha\right)^{1/\alpha}$, i.e., $t \le 0.4 N^{2.5}$ for
$\alpha = 0.8$ which is not a serious limitation actually because our translocation times
scale as $\tau \propto N^{1.5}$ as will be demonstrated below. 

\section{Statistical moments $\left\langle s \right\rangle $ and $\left\langle
s^2 \right\rangle$  vs. time } The recording of statistical moments time dependence,
$\left\langle s(t) \right\rangle = \int_{0}^N s w (s, t) ds$ and $\left\langle s(t)^2
\right\rangle = \int_{0}^N s^2 w (s, t) ds$, is very instructive
(as in the force - free case \cite{Dubb}) for the consistency  check.
Starting from eq. (\ref{Norma}) the calculation of the first moment yields

\begin{eqnarray}
\left\langle s(t) \right\rangle  &=& f {\tilde t} +  2\sqrt{\frac{{\tilde
t}}{\pi}}
\nonumber\\
&\times&\frac{\exp[-f^2 {\tilde t}/4] - \exp [-(f
\tilde{t}-N)^2/4\tilde{t}]}{{\rm erf}[f \sqrt{\tilde{t}}/2] - {\rm
erf}[(f\tilde{t} - N)/2\sqrt{\tilde{t}}]}
\label{First_mom}
\end{eqnarray}
It can easily be shown that in  the  large time limit $\left\langle s
\right\rangle \rightarrow N$.
In the same manner the second moment reads

\begin{eqnarray}
\label{Second_mom}
&&\left\langle s^2(t) \right\rangle  = f^2 {\tilde t}^2 + 2 {\tilde t} + 
2\sqrt{\frac{{\tilde t}}{\pi}}
\\
&\times& \frac{f\tilde{t}\exp[-f^2 {\tilde t}/4]-(f\tilde{t}+N)\exp [-(f
\tilde{t}-N)^2/4\tilde{t}]}{{\rm erf}[f \sqrt{\tilde{t}}/2] - {\rm
erf}[(f\tilde{t} - N)/2\sqrt{\tilde{t}}]}\nonumber
\end{eqnarray}
In eqs. (\ref{First_mom}) and (\ref{Second_mom}) the notations are the same as
in eq. (\ref{Norma}). The detailed check of these relations will be given below.
Here we only note that for large times $1<\tilde{t}<N/f $ the exponential terms in eqs.
(\ref{First_mom}) and (\ref{Second_mom}) vanish so that to a leading order the moments 
vary as $\left\langle s(t) \right\rangle \propto t^{\alpha}$ and $\left\langle 
s(t)^2 \right\rangle \propto t^{2\alpha}$. Again it can be shown that at $t\rightarrow 
\infty$ the moments $\left\langle s(t) \right\rangle$ and $\left\langle s^2 (t)
\right\rangle$ saturate to plateaus which scale like $N$, and $N^2$ respectively, 
as they should.

\section{Scaling arguments} The foregoing theoretical consideration has been  based
on a rigorous mathematical treatment of the FFPE. Before proceeding to the MC -
check of these findings we put forward some simple scaling arguments so as to quantify the
mean translocation time as well as the statistical moments. Let us take the
average external field energy $|\left\langle U (s) \right\rangle| = f N$ as a
natural scaling variable. Then the driven translocation rate scales as
$\tau^{-1}  = \tau_0^{-1} \phi (f N)$, where $\tau_0 \propto
N^{2\nu+2-\gamma_1}$ denotes  the translocation time in the force - free case
\cite{Dubb}. The scaling function $\phi (x)$ behaves in the following way:
$\phi(x \ll 1) \simeq 1$ and $\phi(x\gg 1) \simeq x$ because at $f N \gg 1$ we
could expect that the translocation rate is proportional to the force $f$. As a
result we come to the conclusion that at $f N \gg 1$ the translocation time is
scaled as
\begin{equation}
\tau \propto \frac{1}{f}\:  N^{2\nu + 1 - \gamma_1}
\label{Tau}
\end{equation}
Taking into account the values for $\nu$ and $\gamma_1$ given above we arrive at
the estimations: at $d = 3$ the translocation exponent  $\theta = 2\nu + 1 -
\gamma_1 = 1.496$ and at $d = 2$ the exponent  $\theta = 2\nu + 1 - \gamma_1 =
1.56$. This is pretty close to the estimation given by Kantor\& Kardar
\cite{Kantor2}, $\theta = 1.53$.

If we assume that the behavior of $\left\langle s^2 (t)\right\rangle$  (before
it hits the plateau) follows a power law, $\left\langle s^2 (t) \right\rangle 
\propto (f t)^{\beta}$, then from the correspondence to the scaling law, eq.
(\ref{Tau}), one may estimate $\beta$. Indeed, at the translocation
time $(f \tau)^{\beta} \sim N^2$ and the requirement of correspondence with eq.
(\ref{Tau}) yield $\beta = 2/(2\nu + 1 - \gamma_1)$. This gives $\beta =
1.334$ at $d = 3$.  In the next section we will demonstrate that this power
law is in reality too crude. The observed exponents in the MC simulation (as well 
as in the analytical theory given above) cross over from a smaller value at very short
time to a larger one ($2\alpha \approx 1.6$) at long times in comparison 
with the simple  scaling prediction $\beta = 1.334 \approx 4/3$.

\section{ Monte Carlo data vs. theory}
We have carried out extensive MC
- simulations in order to check the main predictions of the foregoing analytical
theory. We use a dynamic bead-spring model which has been described before
\cite{Milchev}, therefore we only mention the salient features here. Each
chain contains $N$ effective monomers (beads), connected by anharmonic
FENE (finitely extensible nonlinear elastic) springs, and the nonbonded segments
 interact by a Morse potential. An elementary MC move is performed by picking an
effective monomer at random and trying to displace it from its position
to a new one chosen at random. These trial moves are
accepted as new configurations if they pass the standard Metropolis acceptance
test. It is well established that such a MC algorithm, based on local moves,
realizes Rouse model dynamics for the polymer chain. In the course of the
simulation we perform successive runs for chain lengths $N=16, 32,64, 128, 256, 512$
whereby a run starts with a configuration with only few segments on the
trans-side. Each run is stopped, once the entire chain moves to the
trans-side. Complete retracting of the chain back to the cis-side is
prohibited by taking the head monomer larger than the pore diameter. During each 
run we record the translocation time $\tau$, and the
translocation coordinate $s(t)$. Then we average all data over typically $10^4$
runs.
\begin{figure}[th]
\centerline{\includegraphics[scale=0.4,angle=270]{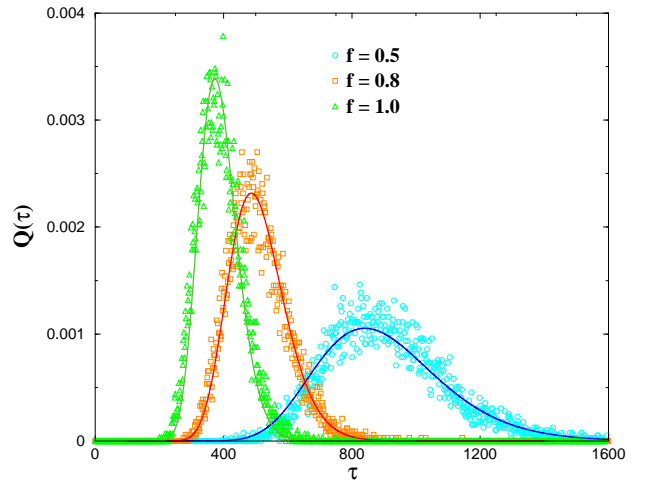}}
\caption{ First passage time distribution functions at $N = 128$ and different
forces as calculated from MC - data (symbols) and the theoretical prediction eq.
(\ref{FPDT_norma}) (solid lines). }
\label{PDF}
%\twocolumn
\end{figure}
In Fig. \ref{PDF} we show the PDF $Q(\tau)$ of a polymer chain with $N=128$ 
for three different values of the drag force, $f=0.5, 0.8\; \mbox{and}\; 1.0$. Although the MC data 
is somewhat scattered, especially for $f=0.5$, the agreement with the analytic
expression,
eq. (\ref{FPDT_norma}) is very good. Since we set the generalized diffusion coefficient 
$K_{\alpha} \equiv 1$ and $\Gamma(1+\alpha) \approx 0.931$ for $\alpha = 0.8$, the comparison
with MC results suggests that a time unit in the FFPE corresponds roughly to $500 MCS$.

\begin{figure}[h]
\centerline{\includegraphics[scale=0.4,angle=270]{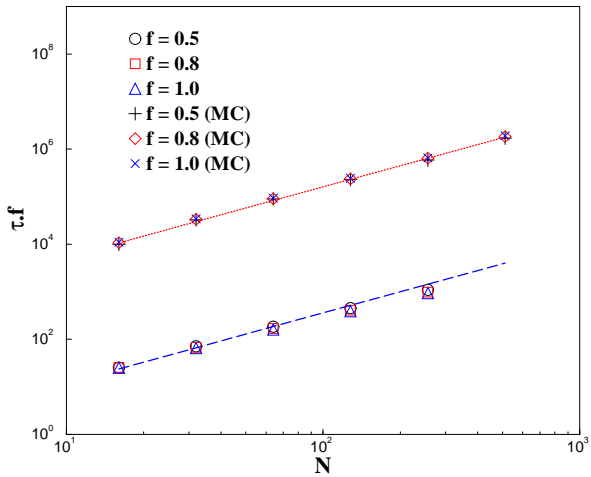}}
\caption{ The average translocation time versus chain lengths. The upper line
represents the results of MC - simulation, the lower refers to the theoretical
prediction obtained by the proper numerical integration of FPTD
eq.(\ref{FPDT_norma}). Both  lines  correspond to a power - law dependence
with exponent $1.5$}
\label{Tau_vs_N}
%\twocolumn
\end{figure}

\begin{figure}[ht]
%\twofigures[scale=0.3]{sketch.eps}{barrier.eps}
\centerline{\includegraphics[scale=0.3,angle=270]{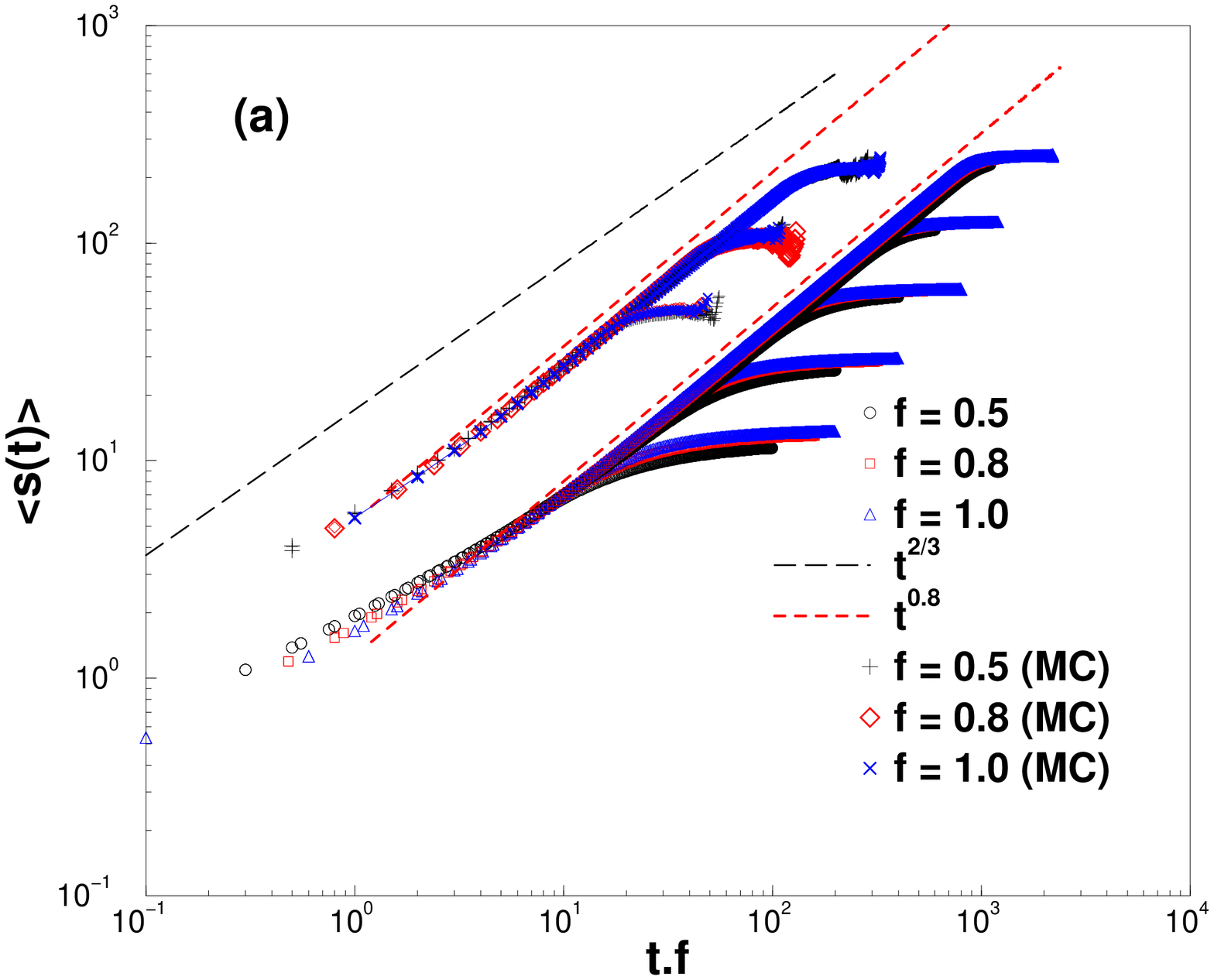}}
\centerline{\includegraphics[scale=0.3,angle=270]{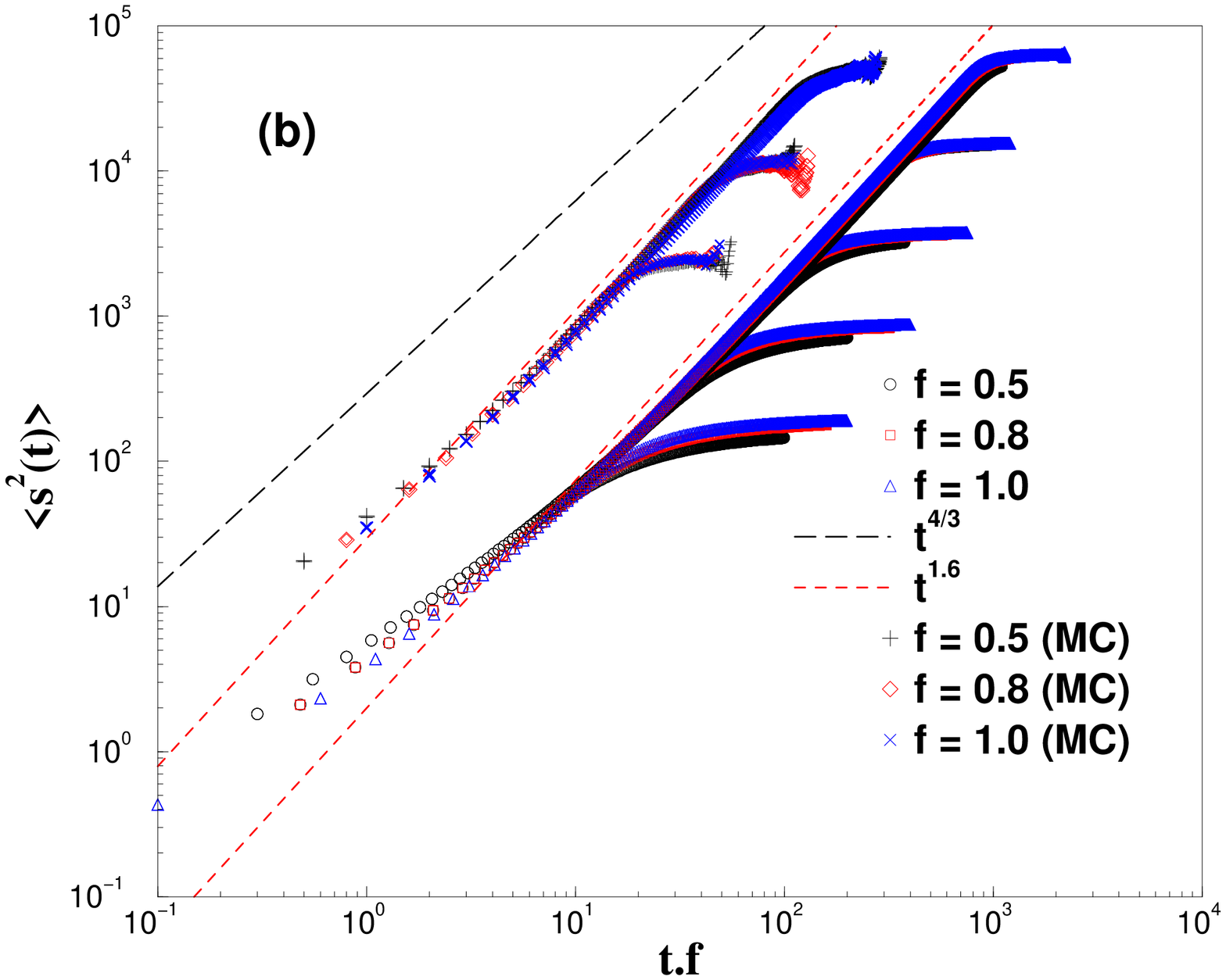}}
\caption{ Statistical moments versus reduced time $t f$ from MC data and from the
analytic results, eqs. (\ref{First_mom}), (\ref{Second_mom}) for chain length 
$16 \le N \le 256$: (a) The first moment 
$\left\langle s(t) \right\rangle$: the slope $\beta/2 = 2/3$ is indicated by a long
dashted line, a short dashed line denotes $\langle s\rangle \propto t^{\alpha}$. 
(b) The second moment $\left\langle s^2(t) \right\rangle$: A long dashed line indicates 
a slope $\beta = 4/3$, a short dashed line denotes 
 $\langle s^2\rangle \propto t^{2\alpha}$.}
\label{First_and_second}
%\twocolumn
\end{figure}

Using the PDF $Q(\tau)$, one may determine the MFPT (or, translocation times) $\tau$ which
are compared in Fig. \ref{Tau_vs_N} for $16<N<512$ and six values of the drag force $f$. 
Evidently, for both theory and simulation the data collapse on master curves
$f\tau \propto
N^{1.5}$, if one scales $\tau$ with the respective force, cf. eq.(\ref{Tau}). It is seen 
that the simulation data is shifted up by a factor of $\approx 500$ which translates the MC
time into conventional time units.
The variation of the moments $\langle s\rangle$, and $\langle s^2\rangle$
is displayed in Fig. \ref{First_and_second}. Again a perfect collapse of the
transients is achieved by scaling the time with the applied force $t \rightarrow
tf$. One can immediately see that the simple scaling prediction $\langle 
s^2\rangle \propto t^{\beta}$ is not perfect:
for $tf < 3$ evidently $\langle s\rangle$ grows with a smaller exponent whereas 
at later times the increase is steeper. As mentioned above, this course is
very well accounted for by eqs. (\ref{First_mom}), (\ref{Second_mom}). Thus, for
$tf \ll 1$ one can readily obtain from eq. (\ref{First_mom}) as a leading term
$\langle s\rangle \propto t^{\alpha/2}$ while for $1 < tf < N$ one has $\langle 
s\rangle \propto t^{\alpha}$. As indicated in Fig. \ref{First_and_second}, the 
observed agreement between theory and computer experiment is remarkable indeed.
Notably this finding suggests that even the presence of
drag force {\em does not} eliminate the anomalous character of the translocation
process as one would intuitively expect. This result resolves thus a problem, raised 
initially by Metzler and Klafter\cite{Metzler2}. The universal exponent $\alpha = 2/(2\nu +
2 - \gamma_1)$ for unbiased threading through a pore is not suppressed by the
drag
force!
One may thus conclude that the measurement of the number of translocated 
segments with time could provide a means for direct observation of anomalous 
diffusion. 
\section{Summary}
By solving the {\em fractional} Fokker-Planck equation for a driven polymer translocation
through a narrow pore and deriving a closed analytic expression for the probability 
distribution function $W(s,t)$ to have the segment $s$ of the chain in the pore at time $t$
we have demonstrated that the translocation process displays all features typical for 
anomalous diffusion. The physical background of this behavior is elucidated by 
scaling considerations. The polymer translocation is considered as a squeezing of 
subsequent chain fragments (folds), each being in local thermodynamic equilibrium, 
through a narrow pore.
This consideration gives rise to an universal scaling exponent for anomalous
diffusion $\alpha=2/(2\nu+2-\gamma_1)$ so that the time $\tau$ needed for a chain 
of $N$ segments to move from {\em cis} to the {\em trans} semispace in the absence of 
drag scales as $\tau \propto N^{2/\alpha}$. The presence of external pulling force 
modifies this relationship to $\tau \propto f^{-1}N^{2\nu+1-\gamma_1}$. This 
principal result of the present investigation is unambiguously confirmed by
calculation of the mean first passage times (the average translocation times) from 
the derived analytic expression for the translocation time distribution function 
$Q(\tau)$ as well as by comparison to the results of extensive Monte Carlo simulations. 
We also show that the growth of the average number of translocated segments $\langle s
\rangle$ with time follows a power law $\langle s(t)\rangle \propto t^{\alpha}$ 
(for relatively long times) which
directly displays the anomalous diffusion exponent $\alpha$. Our analytic data also 
appears to be in perfect agreement with the simulation results in a wide range of
polymer lengths and forces. Thus we have demonstrated that the translocation
dynamics of a driven polymer chain through a narrow pore retains all features of
anomalous diffusion despite the application of external force.

\acknowledgments
The authors greatfully acknowledge the SFB - DFG 625 project for financial
support. A. M. appreciates hospitality during his stay at the Max-Planck Institute 
for Polymer Research in Mainz.

\end{document}